# Graph-based Collaborative Ranking

Bita Shams [a] and Saman Haratizadeh [a]
[a] University of Tehran, Faculty of New Sciences and Technologies
North Kargar Street, Tehran, Iran 1439957131

**Abstract**

Data sparsity, that is a common problem in neighbor-based collaborative filtering domain, usually complicates the process of item recommendation. This problem is more serious in collaborative ranking domain, in which calculating the users' similarities and recommending items are based on ranking data. Some graph-based approaches have been proposed to address the data sparsity problem, but they suffer from two flaws. First, they fail to correctly model the users' priorities, and second, they can't be used when the only available data is a set of ranking instead of rating values.

In this paper, we propose a novel graph-based approach, called GRank, that is designed for collaborative ranking domain. GRank can correctly model users' priorities in a new tripartite graph structure, and analyze it to directly infer a recommendation list. The experimental results show a significant improvement in recommendation quality compared to the state of the art graph-based recommendation algorithms and other collaborative ranking techniques.

## 1. Introduction

Collaborative filtering (CF) techniques are effective algorithms that help people by filtering irrelevant contents and providing personalized recommendation of useful services. These techniques seek to learn models to predict the services that a user will require in the future based on his preferences in the past.

Collaborative-filtering techniques can be categorized into two classes: rating-oriented and ranking-oriented algorithms. The goal of rating-oriented algorithms is to accurately predict a user's ratings and then, recommend the items with the highest predicted rating for him. On the other hand, ranking-oriented approach, called collaborative ranking, seek to directly predict the rankings of items from the viewpoint of a target user, without explicitly predict the ratings. It has been shown that ranking-oriented collaborative filtering approach is sometimes more intuitive and applicable. To see why, notice that, recommendation is naturally a ranking task and what a recommendation algorithm really needs is to improve the quality of Top-k ranking not predicting the rates (N. Liu & Yang, 2008; Y Shi, Karatzoglou, & Baltrunas, 2012; Y Shi, Larson, & Hanjalic, 2010; Yue Shi, Larson, & Hanjalic, 2013). Moreover, in many applications, all we have is a set of implicit feedbacks while no rating data is available and hence, rating based methods can't be used in such a situation. Note that despite rating and other kinds of explicit feedback, that require the user to explicitly assess the items, implicit feedback can be automatically gathered by tracking the user's interactions with the system (e.g. click, buy, like, etc.). Ranking oriented collaborative filtering can be applied in such situations as well.

Neighbor-based collaborative filtering, one of the main classes of collaborative filtering, estimates the ranking/rating of target user based on the behavior of similar users. Despite several researches in this class of algorithms, they still are not able to precisely calculate users' similarities. The reason can be explained by *sparsity problem* which refers to the fact that in recommender systems, users have given feedback to a small proportion of items, and consequently, they rarely have enough common items or pairwise comparisons for estimation of their true similarities/ dissimilarities (Desrosiers & Karypis, 2011). One approach to overcome this issue, is graph-based recommendation that takes advantages of heterogeneous information networks, that are information networks containing different types of nodes and edges, to refine the similarity measures (M. S. Shang, Fu, & Chen, 2008; Z.-K. Zhang, Zhou, & Zhang, 2010; Zhou, Ren, Medo, & Zhang, 2007), expand the neighborhoods, and, directly calculate the closeness of users and items(Chiluka, Andrade, & Pouwelse, 2011; Silva & Zaki, 2013; Xiang et al., 2010; Yao, He, Huang, Cao, & Zhang, 2013).

Graph-based recommendation algorithms represent the relations between users and items as a bipartite graph in which there is a weighted or unweighted link between a user and each item he has rated (Li & Chen, 2013; M. S. Shang et al., 2008; M.-S. Shang, Zhang, Zhou, & Zhang, 2010; Ting, Yan, & Xiang-wei, 2013; Xiang et al., 2010; Z.-K. Zhang et al., 2010; Zhou et al., 2007). Unfortunately, this approach is basically designed for rating/binary feedbacks and has crucial insufficiencies for ranking-oriented class of neighbor-based collaborative filtering.

The first problem is that current graph-based approaches are incompetent to capture the preference order of users. We refer to the example of Fig.1a to illustrate this shortcoming. Mike and Lee have the same preference order for item *A,* and *B*, while Mike, and Martin have completely opposite preference orders on all items. Current graph-based algorithms represent this data as Fig.1b (Sawant, 2013; M. S. Shang et al., 2008). Intuitively, under this graph modeling, most of well-known graph proximity measures (e.g. common neighbors, distance, Katz, and personalized PageRank) will suggest that Mike is much closer to Martin than Lee that is counterintuitive.

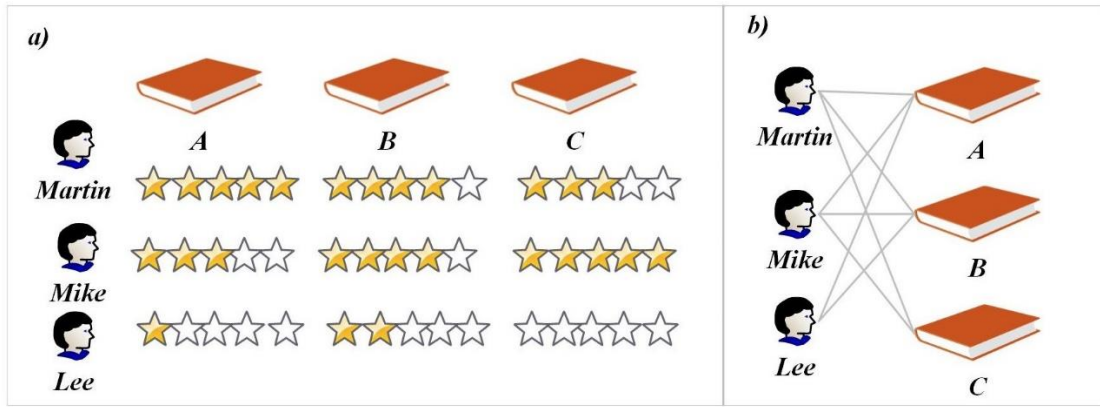

**Fig.1. An example to illustrate incompetence of current graph-based structure to capture preference data gathered in form of rating.**

The second shortcoming of current graph-based approaches have been proposed for binary implicit feedback and that they cannot capture the pairwise preference (i.e. choice context) of user that is generated by different implicit feedbacks. It is clear that the choice context is a valuable piece of information that can be used to improve the recommendation quality. To see how such information is lost when data is modeled by current graph representations, you can observe in the example of Fig.2 that John has preferred item A over B in one session and item B over C in another session, while, Jack has preferred item *B* over *A*. Current graph-based representation of implicit feedbacks, makes a link between the user and those items receiving the positive feedbacks (Chen, Wang, Huang, & Mei, 2012; Xiang et al., 2010; Z.-K. Zhang et al., 2010). Therefore, these algorithms cannot differentiate heterogeneous implicit feedbacks (i.e. buy, click). More importantly, they are not able to clarify the fact that Jack and John disagree when it comes to comparison of items *A* and *B,* as illustrated in Fig.2b.

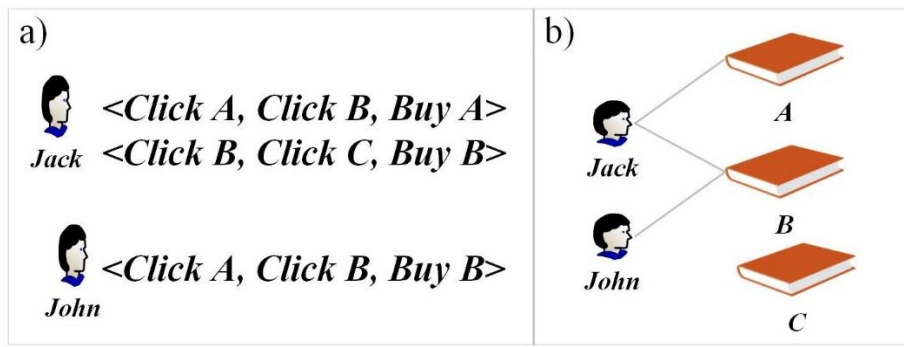

**Fig.2. An example to illustrate incompetence of current graph-based structure to capture the choice context collected through the browsing/ purchasing history.**

This paper presents a novel framework, called GRank, that captures the preference of users using a new Tripartite Preference Graph (TPG) structure that demonstrates the relations between users, items, and pairwise preferences. GRank, also provides a new ranking algorithm, which extends personalized PageRank for top-k recommendation. To the best of our knowledge, this algorithm is the first graph-based approach that is able to capture the preference information provided by implicit feedbacks. Experimental results show higher accuracy of GRank compared to the state of the art collaborative ranking algorithms as well as available graph-based recommendation systems.

The rest of this paper is organized as follows. In Section 2, the related work on graph-based recommendation and collaborative ranking techniques are discussed; then, we present the details of GRank's framework in Section 3. The experimental results are presented and analyzed in Section 4. In Section 5, we discuss how GRank can address some the current shortcomings of collaborative ranking and graph-based recommendation methods. Finally, in Section 6 we conclude and introduce our future works.

## 2. Related Work

The quality of recommendation can be analyzed from many different points of view including accuracy(Koren, Bell, & Volinsky, 2009; Weimer & Karatzoglou, 2007), coverage (Bellogin & Parapar, 2012; Cacheda, Carneiro, Fernández, & Formoso, 2011), diversity (Adomavicius & Kwon, 2012; Said, Kille, Jain, & Albayrak, 2012; Zhou et al., 2010), serendipity(Lu, Chen, Zhang, Yang, & Yu, 2012; Xiao, Che, Miao, & Lu, 2014), uncertainty (M. Zhang, Guo, & Chen, 2015), shilling attack detection(Z. Zhang & Kulkarni, 2014) and scalability (Jiang, Lu, Zhang, & Long, 2011). Although all these aspects are important factors in the success of a recommender system, the recommendation's accuracy is a key element with this regard and a core set of researches have been formed to achieve higher levels of recommendation accuracy. This paper lies in this category, presenting a novel graph-based framework that improves the accuracy of recommendation in the absence of contextual information location when the only available information is the preference data. Contextual information usually refers to the environmental state in which the interaction of the user and the system happens (e.g. time, location, emotion, etc.). On the other hand, the "choice context" reflects the options among which the user makes a choice.

Here, we will review the existing researches related to the main aspects of our proposed algorithm: Collaborative ranking, and Graph-based recommendation.

## 2.1. Collaborative ranking

Collaborative ranking is a class of collaborative filtering algorithms that seeks to predict how a user will rank items. As we mentioned before, despite some similarities, collaborative ranking algorithms are differentiated from rating-oriented collaborative filtering (i.e. collaborative rating) by the fact that collaborative rating algorithms rely on the rating data and try to minimize the rating prediction error, while collaborative ranking algorithms do not depend on rating data. They can use any kind of preference data and try to minimize the rank prediction error.

There is also some similarities between two concepts of collaborative ranking and learning-to-rank problems in the information retrieval domain, as they both try to order entities of on type, e.g. documents/items, for a target entity of another type, e.g. queries/users. However these two kinds of problems are different in practice. In learning-to rank problem there exist a set of explicit common features, such as terms' frequencies among two types of entities, queries and document (Balakrishnan & Chopra, 2012; Fan & Lin, 2013; Y Shi et al., 2010; Volkovs & Zemel, 2012) while there is no such features available or used to relate user and item entities in collaborative ranking problem (Balakrishnan & Chopra, 2012; Y Shi et al., 2010; Volkovs & Zemel, 2012). Because of this important difference between the nature of the problem in these domains, different classes of algorithms have emerged for solving those problems. These approaches can be categorized into two categories: matrix factorization (MFCR) and Neighbor-based algorithms (NCR).

Matrix factorization techniques in collaborative ranking, try to learn representative latent features for an accurate prediction on ranking of items for each user. CofiRank was the first algorithm that uses matrix factorization techniques to optimize a rank-oriented metric(Weimer, Karatzoglou, & Smola, 2008; Weimer & Karatzoglou, 2007). Another technique, ListRank, estimates the Top-1 probabilities to infer a ranking for items. (Y Shi et al., 2010). URM is another model that combines ListRank and probabilistic matrix factorization in order to improve system's accuracy in terms of both ranking and ratings (Yue Shi, Larson, et al., 2013). BoostMF is another matrix factorization approach that sequentially learns a set of weak matrix factorization models based on preference data(Chowdhury, Cai, & Luo, 2015). Bayesian personalized ranking and its variants, try to optimize area under the curve (AUC) for a Bayesian prediction model that is generated based on a set of prediction of pairwise comparisons between relevant and irrelevant items (Lerche & Jannach, 2014; Pan, Zhong, Xu, & Ming, 2015; Rendle, Freudenthaler, Gantner, & Schmidt-thieme, 2009). Recently, some approaches have been proposed that focus on correctly predicting the pairwise preferences for the items with the highest ranks (Christakopoulou & Banerjee, 2015; Dhanjal, Clémençon, & Gaudel, 2015). Climf (Y Shi et al., 2012) and xClimf (Yue Shi, Karatzoglou, Baltrunas, & Larson, 2013) are two other algorithms that exploit matrix factorization techniques to optimize Mean Reciprocal Rank (MRR) of the recommendation list (Y Shi et al., 2012; Yue Shi, Karatzoglou, et al., 2013).

Although the collaborative ranking methods based on matrix factorization consist of a diverse set of algorithms and methods (Weimer et al., 2008; Weimer & Karatzoglou, 2007), the approach of GRank is conceptually different from them as it does not represent the data in a latent feature space. Instead, it models the rank data in the form of a graph structure that enables it to directly estimate the closeness of users and items, based on which it can do the recommendation. So, in a sense, GRank lies in another class of recommendation algorithms called neighbor-based collaborative ranking (NCR). Although this second class of algorithms has its advantages, this approach has remained less investigated, and few successful NCR algorithms have been proposed so far.

EigenRank(N. Liu & Yang, 2008) is the most famous NCR technique that infers a total ranking based on pairwise preferences of users similar to the target user. EigenRank computes users' similarity using Kendall correlation that takes into accounts the agreement and disagreement of users over pairwise comparisons. After estimation of similarities, EigenRank estimates a preference matrix whose elements are a weighted linear combination of neighbors' preferences. Finally, it uses a greedy or Markov-based approach to infer a total ranking over items. To Our knowledge, all of NCR techniques follow the main approach presented by EigenRank with slight modifications. EduRank(AvSegal, Katzir, & Gal, 2014), WSRank (Meng, Li, & Sun, 2011), and Cares(Yang, Wei, Wu, Zhang, & Zhang, 2009) customized EigenRank for different applications. VSRank (Wang, Sun, & Gao, 2014) focuses to improve Kendall similarity measure via considering importance of each pairwise comparison in similarity calculation. However, this approach still suffers from the *sparsity* problem since it still relies on common pairwise comparisons for calculating similarities.

As stated earlier, GRank aims to solve the sparsity problem of neighbor-based collaborative ranking by introducing a novel graph-based approach for modeling and analyzing data. It also differs from the current neighbor-based algorithms as it does not follow the traditional three-step framework, and directly estimates the users' preferences.

### 2.2. Graph-based recommendation approaches

Although there is no graph-based methods designed for collaborative ranking, many recent studies have been conducted in other areas of recommender systems. Here, we will briefly review those algorithms and clarify the main differences between the current work and them.

Graph-based recommendation algorithms are composed of two steps: Constructing a graph representing the data and making recommendations by analyzing the graph. These recommendation algorithms have exploited different types of graphs. However, in all of them, the main component of the graph is the relations between users and those items that have been rated by them. Therefore, the most common approach is constructing a bipartite network where the connections are from one part of the network, users, to the other part, items. Once the bi-partite graph is constructed, several approaches can be used to rank the items using the information from the neighbors of the target user. Approaches like using common neighbors, Katz similarity, diffusion scores and personalized PageRank have been used in this domain (Huang, Li, & Chen, 2005; Z.-K. Zhang et al., 2010)

Recent methods have extended the bi-partite network by adding some layers to it. Some researchers (Xiang et al., 2010) have considered using a session layer to take into account the long-term and the short-term preferences of the user in order to make recommendations in a particular time. Others (Yao et al., 2013) have used different types of nodes in a multi-layer structure to make context-aware recommendation through a random walk in the graph. In (Z.-K. Zhang et al., 2010) a three-layer graph is used to improve recommendation through considering the tags assigned to items by users using a diffusion-based score introduced in (Zhou et al., 2007). In some works (Lee, Park, Kahng, & Lee, 2013; Yu, Ren, Sun, & Gu, 2014) the structure of the network has been revised. They consider a star heterogeneous network, where users and items can be connected to different types of nodes. They use this graph structure to improve the model–based recommendations (Yu et al., 2014) or to make recommendation through improvement of personalized PageRank algorithms in heterogeneous networks. (Lee et al., 2013). We emphasize that none of these algorithms are designed to capture the choice context and preferences of users. Also most of them depend on the contextual information (e.g. time, content, etc.) that does not exist or is not available to the system in all applications and may be expensive to collect.

## 3. GRank: a graph-based framework for collaborative filtering

It has been shown that heterogeneous information networks have strong capabilities to model the relationships among different entities of recommender systems (Cong, 2009; Sun, Han, Yan, & Yu, 2011; Yu et al., 2013, 2014). In this paper, we seek to propose an effective graph approach to ranking-oriented recommender systems, called Graph-based collaborative Ranking, or GRank. In the following, we first define the problem of graph-based collaborative ranking and its purposes. Then, we present some definitions that are needed to understand the algorithm. Next, we introduce a novel heterogeneous graph structure, called tripartite preference graph (TPG) that embeds different kinds of relations among users, preferences and items in an aggregated structure. Finally, we suggest an efficient algorithm to exploit TPG in order to rank items for each target user.

### 3.1. Problem definition

From the collaborative ranking perspective, recommender systems can be represented by the set of users $U = \{u_1, \dots, u_M\}$, set of items $I = \{i_1, \dots, i_N\}$ and the observation set $O = \{<u, i, j>\}$ that is the set of preferences occasionally stated by users. Generally, we define the observation $o = <u, i, j>$ where $u \in U, i \in I$ and $j \in I$ denoting that the user $u$ has preferred i over j. For simplicity, we call the first item, the desirable item and the second one the undesirable item.

Note that pairwise comparison is a general form of ranking data, and, all kinds of preferences (e.g. rating, browsing history) can be converted to a set of pairwise comparisons using the following rules:

*Rule 1. Let L be a rating matrix in which $L_{ui}$ represents the rating of user u for item i. The preference observation set can be obtained by $O = \{<u, i, j> | L_{ui} \neq 0, L_{uj} \neq 0, L_{ui} > L_{uj}\}$*

*Rule 2. Let L be the matrix of positive feedbacks (e.g. Like) in which a non-zero element $L_{ui}$ represents that user u likes item i. Similarly, Let D be the matrix of negative feedbacks (e.g. dislike) in which a non-zero element $D_{ui}$ represents the user u dislikes item i. The preference observation set can be obtained by $O = \{<u, i, j> | L_{ui} \neq 0, D_{uj} \neq 0\}$*

*Rule 3. Let W be the set of sessions that are defined as $W = \{w_1, \dots w_{|w|}\}$, we can create the observation set as*

$$O = \{<u, i, j> | \exists w \in W \: u = w.u, i \in w.B, j \in w.C\}$$

*Where w.u is the user in the session w, w.B is the set of items bought in the session w and w.C is the set of items clicked but not bought in session w*

Given a preference dataset $O$, a graph-based framework will face a key question that is how to model the information available from the preference data set in the graph structure. To answer this question, we first categorize some information that an effective graph modeling of rank-oriented recommender system ought to capture:

- **Users' similarities' in terms of priorities:** Two users of a ranking-oriented recommender system, are assumed to be similar when they have either similar opinions about certain pairwise comparisons(N. Liu & Yang, 2008; Wang et al., 2014) or prefer a particular item *A* in some comparisons even if the items over which it has been preferred are different for those two users (Meng et al., 2011). A well-organized graph model of ranking data should reflect both type of similarities among users.

- **Correlation among comparisons:** Correlated pairwise comparisons are those preferences that are similarly voted by users. These comparisons should be simply discovered by analyzing the graph representation of the data.

- **Items' similarities:** Similar items are those that are similarly favored/disfavored by a group of similar users. An effective graph modeling should clearly reflect the closeness of these items.

- **Prediction of users' priorities:** The ultimate goal of ranking-oriented recommender systems is to infer the total ranking of target user over unseen items and recommend the top-k items. Consequently, graph representation of these systems is responsible for the efficient and effective recommendation to the target user, and ideally, a graph representation of rank data may be used to directly predict the rankings.

### 3.2. Graph Construction

Before we can proceed to explain how GRank constructs and exploits a graph structure based on the preference dataset, we need to define some basic concepts:

***Definition. 1***. *A pairwise preference p is a tuple $< i, j >$ denoting the preference of i over j. We call i as the desirable item in p, represented by p.d, and j as the undesirable item in p represented by p.u. The pairwise preference set P is formally defined as $P = \{< i, j > | i \in I, j \in I, i \neq j\}$.*

***Definition. 2***. *A user may have some certain preference over two items or not. The agreement function $f: U \times P \rightarrow \{0,1\}$ indicates whether the user $u_i$ agrees with the preference $p_j$ or not and is defined as:*

$$f(u_i, p_j) = \begin{cases} 1, & < u_i, p_j.d, p_j.u > \in O \\ 0, & O.w \end{cases}$$

*Where $u_i \in U, p_j \in P$ and O is the observation set of preferences, as defined in section 3.1.*

***Definition.3.*** *Abstractly, each item has two sides: the desirable side and the undesirable side. We define the items' desirability set as $I_d = \{i_d | i \in I\}$ where $i_d$ represents the desirable side of item i. Also the items' undesirability set is defined as $I_u = \{i_u | i \in I\}$ where $i_u$ represents the undesirable side of item i. We also define the representative set as $= I_d \cup I_u$, that will contain two elements for each item, one for each side of it.*

***Definition.4.*** *The support function $s: P \times R \to \{0,1\}$ indicates whether a preference p supports the representative r or not. Formally, we define s as*

$$s(p,r) = \begin{cases} 1, & p.u = i \text{ and } r = i_u \\ 1, & p.d = i \text{ and } r = i_d \\ 0, & O.W \end{cases}$$

*Where $p \in P$ and $r \in R$.*

Based on these definitions and concepts we can now explain how GRank models the preference data using a structure called Tripartite Preference Graph. Formally, *Tripartite Preference Graph (TPG)* is a tripartite graph $TPG(U \cup P \cup R, E_{UP} \cup E_{PR})$, where $U$ is the set of users, $P$ is the set of pairwise preferences and $R$ is the set of representatives. $E_{UP} = \{(u,p) | f(u,p) = 1, u \in U, p \in P\}$, is the set of edges between the nodes in $U$ and $P$, and $E_{PR} = \{(p,r) | s(p,r) = 1, p \in P, r \in R\}$, is the set of edges connecting the nodes in $P$ to the nodes in $R$.

More clearly, TPG contains three layers, each containing a different type of nodes:

- **Users:** TPG's first layer contains one node for each user.

- **Preferences:** The preference layer consist the nodes corresponding to the set of all possible pairwise preferences $p \in P$. Foe simplicity, the corresponding node to each preference $p = <i,j>$ is labeled in the form of ($"i > j"$) that clearly demonstrates the preference of $i$ over $j$.

- **Representatives:** The representative layer contains the set of both undesirable and undesirable representative of items represented by $i_d$ and $i_u$, respectively

TPG also contains two types of links:

- **User-Preference links:** $E_{UP}$ is the set of edges that connects each user $u$ to his stated preference. More clearly, For each preference data $< u, A, B > \in O$, there is a link between user $u$ and the preference node labeled by "$A > B$"

- **Preference-representative links:** $E_{PR}$ is the set of links that connects each preference to the representatives that it supports. For instance, a preference node labeled by "$A > B$" is connected to the nodes corresponding to desirable case of $A$, i.e. $A_d$ and undesirable case of B i.e. $B_u$. This links are used to model the fact that an "$A > B$" preference, implicitly supports item desirable side of "$A$" and undesirable side of "$B$".

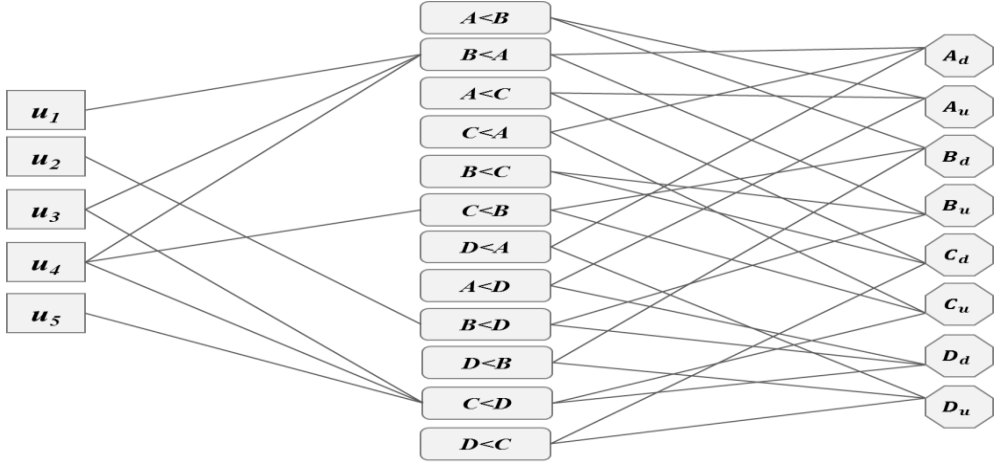

**Fig. 3. An example: A TPG constructed from a system containing 5 users, 4 items, and, 9 assigned pairwise preference.**

As mentioned before, TPG is a tripartite graph in which the preference layer connects to both other layers: the user layer and the item layer. Traversing TPG through different types of paths reveals different types of information in a ranking-oriented recommender systems, some examples of which are presented in Table.1.

**Table 1. The meta-paths and their semantics in TPG**

| Meta-path[a] | Semantic |
|---|---|
| $U - P - U$ | Users' similarities in terms of pairwise preference |
| $U - P - I_d - P - U$ | Agreement of users over desirability/ undesirability of an item over different items. |
| $U - P - I_u - P - U$ | |
| $P - U - P$ | Correlation between pairwise comparisons |
| $I_d - P - U - P - I_d$ | Direct relations between items; items that are simultaneously favored/ disfavored by users |
| $I_u - P - U - P - I_u$ | |
| $I_u - P - U - P - I_d$ | Indirect relationships between items. Items that are contrarily ranked by user. |
| $I_d - P - U - P - I_u$ | |

The pseudo code for constructing TPG is presented in Algorithm.1. It first generates $M$ nodes corresponding to $M$ users and $2N$ nodes for desirable and undesirable case of $N$ items. Then, it generates $N(N - 1)$ nodes for each possible preference data. Each preference data has desirable and an undesirable sides that are represented by *p.d and p.u*, respectively. The next step, is to add links between users and preferences. It scans the preference database and for each $<user, item1, item2>$ triple, it adds a link between the corresponding nodes in the user and the preference layers. In more details, *getUserNode(u)* returns the corresponding node to user *u while getPreferenceNode(i,j) returns* the node *p* representing the preference of *i* over *j*. After that, a link is created between the node *u* and *p*. Finally, the algorithm scans through the preference nodes in which, a preference node p states that "p.d > p.u". For such a node, the algorithm finds *the* corresponding representative nodes using *getDesirableNode(p.d)* and *getUnDesirableNode(p.u)* and inserts an edge between *p* and each of those representatives.

**Algorithm 1. Construction of Tripartite Preference Graph (TPG)**

**Input:** Set of users U, Set of items I, Observation set of preference (O)
**Output:** Tripartite graph (*G*)

Initialize a graph G
*//Initializing user layer*
For each item $u \in U$
  Create a node *u* in user layer

*//Initializing representative layer*
For each item $i \in I$
  Create a desirable-node $i_d$ in the representative layer
  Create an undesirable-node $i_u$ in representative layer

*//Initializing preference layer*
For each item $i \in I$
  For each item $j \in I$
    If $(i \neq j)$
      Create a node *p* in preference layer
      p.d = i;
      p.u=j;

*// Connecting user and preference layer*
For each o:(u,i,j) $\in O$
  u = G. GetUserNode (u);
  p = G. GetPreferenceNode (i ,j);
  Connect node *u* to node *p*.

*//Connecting preference and item layer*
For each node v $\in$ preference-layer
  $r_d$= G.GetDesirableNode (p.d)
  $r_u$= G.GetUnDesirableNode (p.u)
  Make a link between *p* and $r_d$

> Make a link between p and $r_u$.

**Example.1.** Fig.4 illustrates how TPG can reflect the preference data mentioned in Fig.1 and Fig.2. As shown in Fig.4b, TPG clearly indicates that Mike and Martin (in Fig.1a and Fig.4a) have not the same preference as they neither share any neighbors nor longer paths to each other in TPG. On the other hand, Mike and Lee share one common neighbor that denotes their agreement over comparison of A and B. The same holds for Jack and John in the example of Fig.4c where TPG representation (Fig.4d) can evidently reflect that they do not have the same opinion over the comparison of A and B. Note that current graph-based approaches are not capable to model users' preference in these samples as mentioned in Fig.1b and Fig.2b.

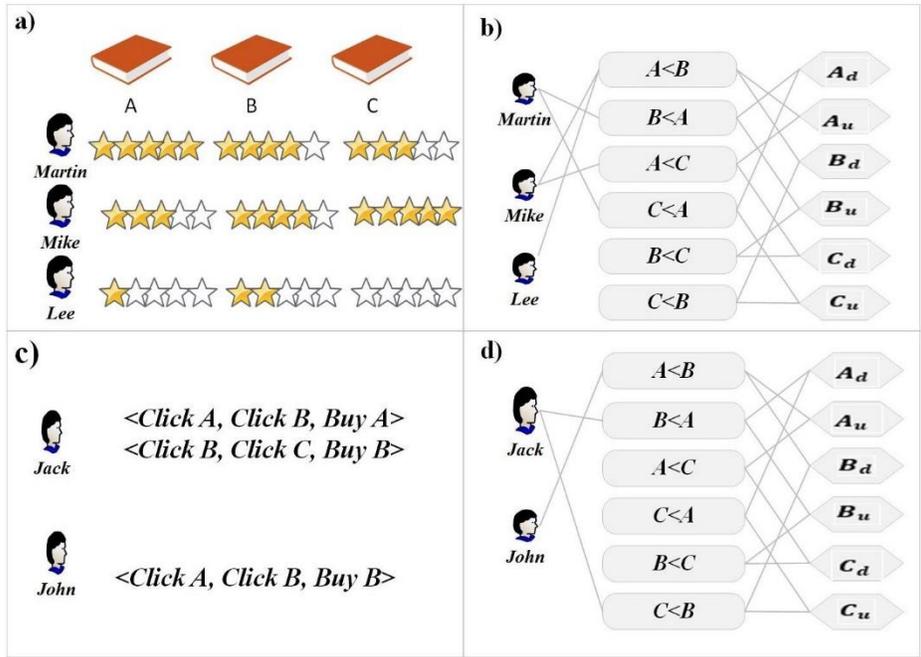

**Fig.4. TPG representation of examples in illustrated in Fig.1, and, Fig.2, respectively.**

### 3.3. Top-k recommendation using TPG graph

It is a common assumption in recommender systems that users are interested in items that are preferred by their neighbors or are similar to their favorite items. As mentioned before, TPG provides an affluent platform to determine the users' similarity and items' relations. GRank exploits the closeness of users to the desirable/undesirable representatives to estimate how much the target user likes/dislikes a specific item. In other words, given a target user $u$ and a TPG describing the observation set, GRank defines a function $GR: U \times I \to \mathbb{R}$ for predicting the goodness of each unseen item i for each user *u based on the closeness of the node u to the* desirable and undesirable representative nodes of item i. Then GRank recommends to u the items with the highest *GR* values calculated for u.

The desirability/undesirability of an item *i* for user *u* can be estimated based on two general types of paths: Desirability and Undesirability paths. Desirability paths are in form of $< u, v_1, v_2, \ldots, v_m, i_d >$ and show the closeness of a target user *u* to the desirable case of an item *i*. Similarly, undesirability paths are in form of $< u, v_1, v_2, \ldots, v_m, i_u >$ and depict the closeness of the target user *u* and the undesirable case of item *i*.

Intuitively, there are a large number of desirability/undesirability paths between the target user *u* and an item *i*. Therefore, some form of proximity measure is required to compare the number of desirability/undesirability paths between the target user and items.

Recently, some proximity measures have been proposed for analyzing heterogeneous networks. Unfortunately, these methods are heavily dependent to the definition of meta paths and require to get the importance weight of different meta-paths as an input parameter to calculate the proximity among nodes(Lee et al., 2013; Sun et al., 2011). On the other hand, a general approach for measuring nodes' proximity in a network, is personalized PageRank, or PPR (Page, Brin, Motwani, & Winograd, 1999) that is acknowledged as one of the most effective measures that ranks nodes based on their reachability from a certain set of nodes in a network. It gives high scores to items that are closer to the target user regarding a wide range of graph properties such as distance or number of paths between them (Lee et al., 2013). GRank defines a measure based on PPR to calculate proximities, but before introducing the measure, first we briefly review the concept of PPR.

Formally, the personalized PageRank of a node indicates the probability that a random walker, with a given skewed restarting distribution, will jump to that node. PPR can be considered as a Markov process with restart, and, is defined by Eq.1

$$\text{PPR}(t) = \alpha \cdot T \cdot \text{PPR}(t-1) + (1-\alpha) \cdot PV \tag{1}$$

where PPR(t) denotes the rank vector at the $t-th$ iteration, $T$ is the transition matrix, $\alpha$ is the damping factor and "$PV$" is the user-specific personalized vector. In most of applications $\alpha$ is set to 0.85.

To calculate closeness of some graph nodes, such as item nodes, to a particular node, like the target user's node, personalized PageRank needs to define a personalized vector *PV* as in Eq.2

$$PV(j) = \begin{cases} 1, & j = u \\ 0, & otherwise \end{cases} \quad (2)$$

where $u$ is the target node.

Given a graph $G(V, E)$ with the set of all nodes $V$ and the set of all edges $E$, each element of the transition matrix is obtained from Eq.3

$$T_{ij} = \begin{cases} \frac{1}{d_i}, & (i,j) \in E \\ 0, & otherwise \end{cases} \quad (3)$$

Where $d_i$ is the degree of the $i-th$ node.

As aforementioned, PPR can be used to find the closeness of each node to the target user's node in TPG. More clearly, personalized PageRank of the target users in TPG, estimates the probability that a random walker, starting from the target user, will follow a path to the desirable and undesirable representative of each item. GRank defines the goodness of an item $i$ for the target user based on PPR of the desirable and undesirable cases of $i$, as in Eq. 4

$$GR(i) = \frac{PPR(i_d)}{PPR(i_d) + PPR(i_u)} \quad (4)$$

One can expect that $GR(i)$ gives top scores to those items for which personalized PageRank of their desirable case is much higher than that of their undesirable case. To make a recommendation to the target user, items are sorted according to their GR values, and the top-k items are suggested.

Algorithm 2 summarizes the GRank's approach for top-k recommendation. GRank requires to calculate the personalized PageRank of nodes for each target user $u$. For this purpose, it first defines the transition matrix of TPG and the personalization vector for u, using Eq.3 and Eq.2, respectively. Then, GRank randomly initializes the PPR values and then normalizes them to their summation. Next, it updates PPR value using Eq.1 and iterates until convergence. After that, GR values are calculated for each item using Eq.4. Finally, the items are sorted based on their GR values and the top-k items are recommended to the target user $u$

**Algorithm 2. Top-K recommendation on TPG**

| |
|---|
| **Input:** Tripartite graph ($G$), Target user $u$, *number of recommended items K, set of Items I* <br> **Output:** The best *k items.* |
| |

```
Initialize transition matrix T through Eq.3
Initialize personalized vector PV through Eq.2
Randomly initialize PPR₀
PPR₀ = PPR₀/sum(PPR₀ )
t=1
Repeat until convergence
        PPRₜ  =  α · T · PPRₜ₋₁ +  (1 − α) · PV
        t=t+1;

For each item i ∈ I
        Calculate GR value for items through Eq.4
Put items in descending  order of their GR values in list L
Return as the recommendation list the first k items in L
```

### 3.4. Computational Complexity

GRank is composed of two phases: Graph construction and recommendation task. Assume that the $M$ is the number of users, $N$ is the number of items and $S$ is the number of total pairwise preference assigned by all users. Clearly to total number of possible pairwise comparison would be $N(N − 1)$.

TPG contains $N(N − 1) + M + 2N$ vertices and $S + 2N(N − 1)$ edges. The time complexity of constructing the graph would depend on the implementation approach. If we use adjacency lists, then it would have a time complexity of $O(N^2 + M + N + S + 2N^2) = O(N^2 + M + S)$. . Additionally, we often know that $S = cN^2$ where c is a small constant (e.g. 2.48 for MovieLense100K). So, the time needed for graph construction phase is in $O(M + N^2)$.

The time complexity of recommendation task is equal to that of personalized PageRank (PPR) computation. Computational complexity of personalized PageRank is $O(tE)$ where $E$, is the number of graphs' edges and $t$ is the number of iterations needed before personalized PageRank converges. In TPG, we have $(S + 2N^2 − 2N)$ edges, and as we mentioned, we expect that $S = cN^2$, for a small constant $c$. So the time complexity of a recommendation in GRank, is expected to be $O(tN^2)$. Note that $t$ is a small number for sparse graphs such as TPG. In our experiments, it does not exceed 20

Note that recommendation by GRank has a better computational complexity than EigenRank, the most acknowledged memory-based CR. EigenRank's computational complexity is in the order of $O(MN^2 + KN^2 + N^2)$ where $O(MN^2)$ is for calculating the similarity between the target user and all other users, $O(KN^2)$ is for estimating the preference matrix, and, $O(N^2)$ for inferring the total ranking.

# 4. Experimental settings and results

We have conducted a series of experiments for evaluating GRank algorithm. Here, we will first give a detailed description of the experimental protocol. Then, we will analyze the ranking quality and scalability of GRank.

## 4.1. Experimental setting

### 4.1.1 Dataset

We conducted all experiments on two publicly available datasets that are widely used in related work (Fan & Lin, 2013; Rendle et al., 2009; Yue Shi, Larson, et al., 2013; Volkovs & Zemel, 2012; Wang et al., 2014): Both data sets have been generated by Movielens group , but contains different number of users, items, and, ratings. The first data set, Movielens-100K, consists of 100,000 ratings (scale 1-5) assigned by 943 users to a collection of 1,682 movies. The second dataset, Movielens-1M, is composed of 3,952 movies rated by 6, 040 users. There are one million ratings in this dataset. Since GRank is designed for using pairwise preference data, we converted the rating information into a set of pairwise comparisons using Rule 1: we created a preference instance of data <u, item#1, item#2) if *item#1* has been rated higher than *item#2* by user u.

### 4.1.2 Evaluation methodology

In our experiments we followed a standard protocol widely used in related work (Balakrishnan & Chopra, 2012; Fan & Lin, 2013; J. Liu, Wu, Xiong, & Liu, 2014; Rendle et al., 2009; Y Shi et al., 2012; Volkovs & Zemel, 2012). We analyzed the effectiveness of our algorithm under different conditions of user profiles, regarding to the number of user's ratings; for each user, a fixed number $T$ of ratings was randomly sampled and placed in the training set, and the remaining ratings went to the test set. Our experiments involve $T = 20, 30, 40, 50$ items. For each $T$, we make sure that we can compare algorithms on at least 10 rated items per user in the test set. Therefore, the users with respectively less than 30, 40, 50, and, 60 items are dropped from both train and test sets. We generated 5 variants of both data sets via random sampling and the average performance on all variants of the test set is reported.

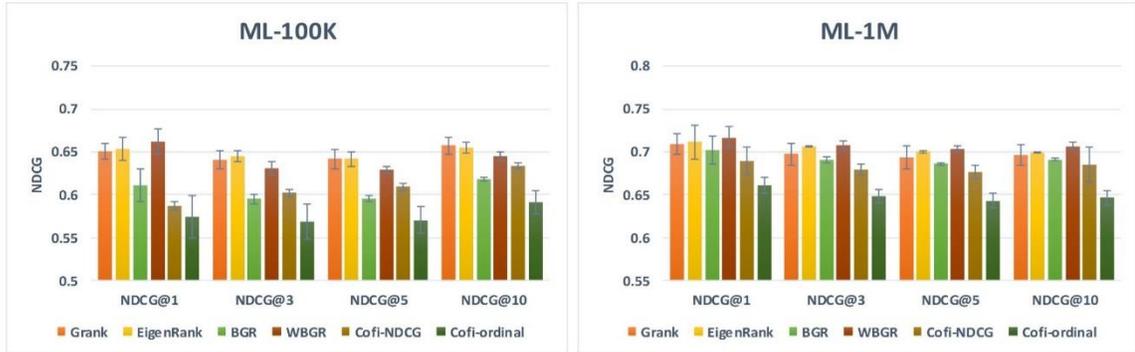

**Fig. 5. Performance comparison of algorithms in terms of NDCG where T=20.**

### 4.1.3 Baseline algorithms

Since GRank makes a connection between neighbor-based collaborative ranking and graph-based recommendation, we have compared its performance against state of the art approaches of these two classes of methods. For further analysis, we also compare our algorithm to Cofi-Rank, the state of the art matrix factorization technique that is able to do recommendation based on pairwise preference data. These algorithms are briefly described in the following:

- **CofiRank:** CofiRank (Weimer & Karatzoglou, 2007) is one of the state of the art MFCR techniques that extracts latent representations in order to optimize a structured loss function. CofiRank has several extensions. In our experiments, we used CofiRank-Ordinal and CofiRank-NDCG as our baseline algorithms. (Weimer et al., 2008) . CofiRank-Ordinal minimizes the number of discordant preferences in the predicted ranking list while CofiRank-NDCG triers to maximize the NDCG. We used the publicly available code for CofiRank and adopted the optimal values for its parameters as suggested in (Weimer et al., 2008).

- **EigenRank:** EigenRank(N. Liu & Yang, 2008) is another famous algorithm in the family of NCR techniques. We have implemented the random-walk version of EigenRank using neighborhood sizes of 100 and $\varepsilon = 0.85$, that have been reported to be the best parameter values for the algorithm (N. Liu & Yang, 2008).

- **Graph-based recommendation:** Also, we compare GRank with a graph-based recommendation algorithm that exploits a bi-partite graph structure to model the user-item interactions (such as Fig.1). Then, a random-walk with restart is used to rank items in the bi-partite graph(Chiluka et al., 2011). We have implemented two versions of this algorithm, abbreviated by BGR and WBGR, which links the users to those items rated by him with un-weighted and weighted links, respectively. In the weighted version, the weight of the edge between user $u$ and item $i$ is set equal to the rating of $u$ to $i$.

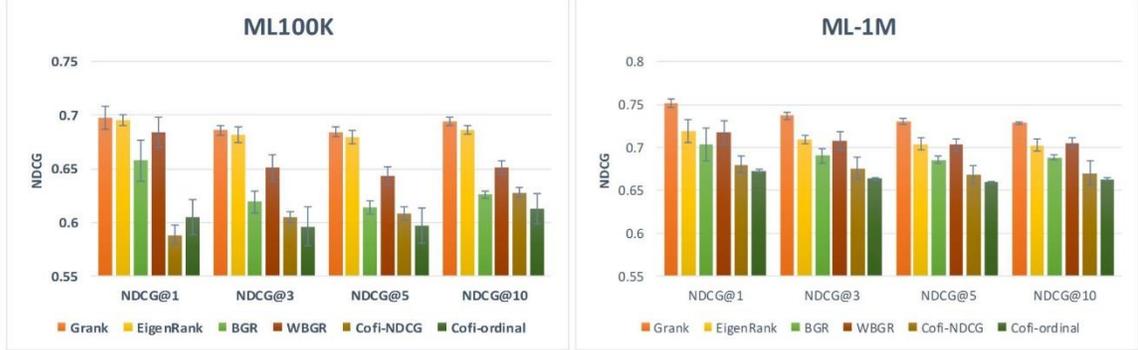

**Fig. 6. Performance comparison of algorithms in terms of NDCG where T=30**

### 4.2. Results

#### 4.2.1 Accuracy

Following the standard evaluation strategy applied to recommender systems, we assessed the recommendation performance of the models by comparing the quality of their top-k suggestions.

Normalized Discounted Cumulative Gain (NDCG) is an evaluation metric that is widely used for assessment of CR techniques. The definition of NDCG at the top-K suggestions for a user $u$ can be given as:

$$NDCG@K = \frac{1}{\alpha_u} \sum_{i=1}^{K} \frac{2^{r_i^u} - 1}{\log(i+1)} \tag{5}$$

Where $K$ is the length of the recommendation list, $r_i^u$ is the rating given by user $u$ to the $i$-th item in the recommendation list, and $\alpha_u$ is the normalization factor to ensure that NDCG of the ideal recommendation for $u$ is equal to 1. In this paper, we report the recommendation performance by NDCG@3, NDCG@5 and NDCG@10, averaged across all users.

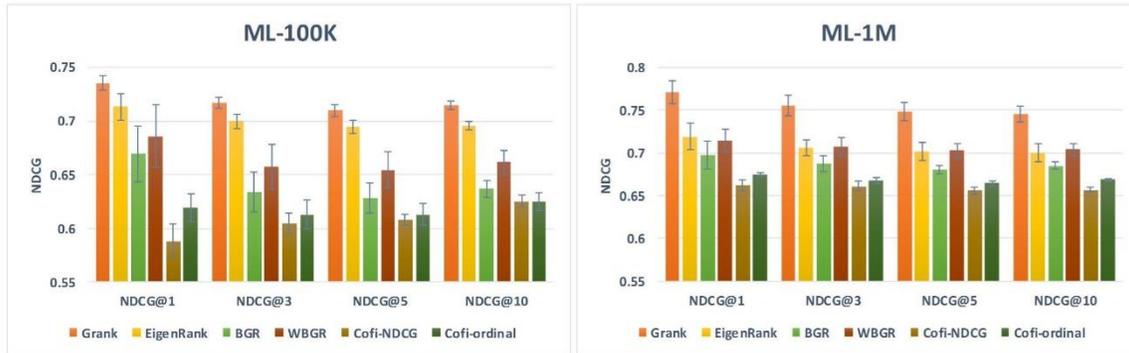

**Fig 7. Performance comparison of algorithms in terms of NDCG where T=40.**

The performance of the algorithms on ML-100k and ML-1M are shown in Fig.5 Fig.6, Fig.7, and, Fig.8. In most of experiments, GRank outperforms other algorithms. The results are reported based on a set of experiments on sufficiently large samples of data, and different algorithms are tested on common sets of samples. So, to test if the differences between the performance of GRank and other algorithms are significant, we can conduct a set of paired t-tests on the results. (Fouss, Pirotte, Renders, & Saerens, 2007; Guo, Zhang, & Yorke-Smith, 2015; Kim & Ahn, 2008; Shani & Gunawardana, 2011; Ting-Peng, Hung-Jen, & Yi-Cheng, 2006; Zhen et al., 2009).

Given $\{x_1, \dots, x_N\}$ be the set of samples derived from the dataset, let $\mu_d$ and $\sigma_d$ as the average and variance of differences $d_i = a_i - b_i$ where $a_i$ and $b_i$ denote the performance of two approaches *A* and *B* on the *i-th* sample. To determine whether A significantly outperforms *B* or not, we consider the null hypothesis is $\mu_d = 0$ whereas the alternative hypothesis is $\mu_d \neq 0$. The null hypothesis is rejected in favor of the alternative hypothesis if the p-value, obtained by the t-statistic $t = \frac{\mu_d}{\sigma_d/\sqrt{N}}$, is below than the significance threshold (e.g. 0.01). Table 2 shows the p-values indicating the significant outperformance of GRank w.r.t other algorithms.

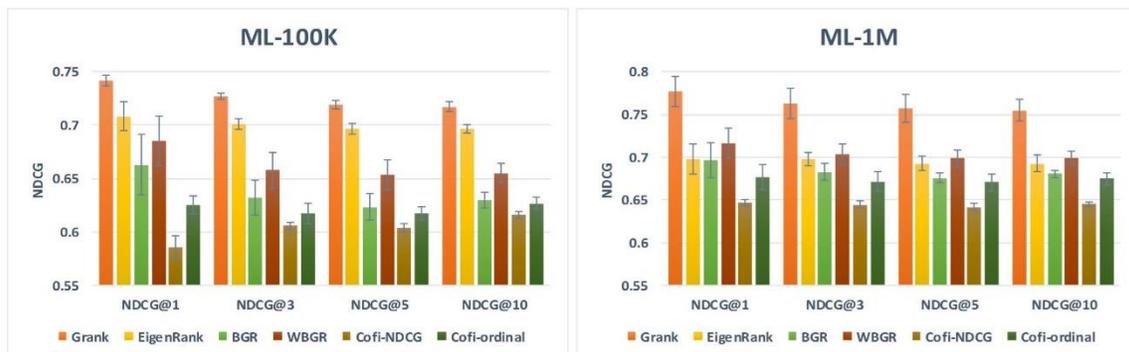

**Fig 8. Performance comparison of algorithms in terms of NDCG where T=50.**

The experimental results can be summarized as below:

- GRank significantly outperforms all algorithms in the majority of evaluation conditions. Yet, EigenRank and WBGR show an improvement of less than 1% in ML-100k and Ml-1M where T=20.

- The performance of GRank is up to 6% and 8% better than WBGR and BGR, respectively. This result reemphasizes the importance of capturing users' preference and their choice context for recommendation. It is worth noting that this feature will be more renowned while increasing T. In case of small number of training data, users rarely have common items that indicate their similarity.

- GRank improves EigenRank about 1%-4% in ML-100K, while, its performance is up to 8% better than EigenRank in ML-1M. This results can be explained by the natural problem of Kendall correlation measure with sparse data. At fixed *T*, the number of items in ML-1M is two times more than ML-100K, and consequently is two times sparser, Therefore, EigenRank faces more difficulties to effectively calculate similarities in ML-1M and so wrong users may be picked by EigenRank as neighbors of a target user. On the other hand, GRank that does not rely on direct similarity calculation among users, can handle such sparse data sets quite well.

Table 2. P-values obtained for the paired t-test under each evaluation condition. P<0.01 indicates the significant out performance of GRank w.r.t other algorithm

| | | ML-100K | | | | | ML-1m | | | | |
|---|---|---|---|---|---|---|---|---|---|---|---|
| $T^a$ | $K^b$ | Eigen Rank | BGR | WBGR | Cofi-NDCG | Cofi-Ordinal | Eigen Rank | BGR | WBGR | Cofi-NDCG | Cofi-Ordinal |
| 20 | 1 | 0.5396 | 0.038 | 0.3653 | 0.0003 | 0.0020 | 0.8310 | 0.7465 | 0.5417 | 0.0959 | 0.0006 |
| 30 | 1 | 0.6952 | 0.0040 | 0.0214 | 0.0002 | 0.0012 | 0.0100 | 0.0095 | 0.0071 | 0.0035 | 0.0000 |
| 40 | 1 | 0.0131 | 0.0040 | 0.0143 | 0.0002 | 0.0001 | 0.0002 | 0.0000 | 0.0003 | 0.0005 | 0.0007 |
| 50 | 1 | 0.0112 | 0.0040 | 0.0072 | 0.0000 | 0.0000 | 0.0051 | 0.0013 | 0.0015 | 0.001 | 0.0009 |
| 20 | 3 | 0.2577 | 0.0020 | 0.3033 | 0.0044 | 0.0005 | 0.2033 | 0.5991 | 0.3539 | 0.0481 | 0.0001 |
| 30 | 3 | 0.2908 | 0.0003 | 0.0041 | 0.0000 | 0.0007 | 0.0008 | 0.0007 | 0.0064 | 0.0073 | 0.0000 |
| 40 | 3 | 0.0069 | 0.0010 | 0.0064 | 0.0000 | 0.0000 | 0.0000 | 0.0000 | 0.0001 | 0.0003 | 0.0007 |
| 50 | 3 | 0.0014 | 0.0002 | 0.0009 | 0.0000 | 0.0000 | 0.0015 | 0.0006 | 0.0007 | 0.0008 | 0.0010 |
| 20 | 5 | 0.9047 | 0.0003 | 0.1455 | 0.0099 | 0.0001 | 0.2910 | 0.5021 | 0.2865 | 0.0321 | 0.0003 |
| 30 | 5 | 0.1842 | 0.0000 | 0.0014 | 0.0000 | 0.0004 | 0.0021 | 0.0002 | 0.002 | 0.0039 | 0.0000 |
| 40 | 5 | 0.0015 | 0.0007 | 0.0051 | 0.0000 | 0.0000 | 0.0000 | 0.0004 | 0.0002 | 0.0003 | 0.0007 |
| 50 | 5 | 0.0060 | 0.0000 | 0.0005 | 0.0000 | 0.0000 | 0.0007 | 0.001 | 0.0008 | 0.0005 | 0.0009 |
| 20 | 10 | 0.3772 | 0.0009 | 0.1446 | 0.0139 | 0.0000 | 0.5587 | 0.6757 | 0.2618 | 0.2020 | 0.0002 |
| 30 | 10 | 0.0070 | 0.0000 | 0.0007 | 0.0000 | 0.0002 | 0.0063 | 0.0000 | 0.0039 | 0.0069 | 0.0000 |

| | | | | | | | | | | |
|---|---|---|---|---|---|---|---|---|---|---|
| 40 | 10 | 0.0005 | 0.0000 | 0.0010 | 0.0000 | 0.0000 | 0.0000 | 0.0003 | 0.0000 | 0.0003 | 0.0005 |
| 50 | 10 | 0.0046 | 0.0000 | 0.0000 | 0.0000 | 0.0000 | 0.0000 | 0.0006 | 0.0004 | 0.0004 | 0.0005 |

### 4.2.2 Scalability

The final experiment investigates the scalability of GRank that, as mentioned in section 3.4, depends on three factors: number of users ($M$), number of items ($N$), and number of assigned preferences ($S$). To investigate the scalability of GRank under each evaluation condition, that is the number of ratings available for each user in the training set (See Section 4.1.2), we measured the running time by altering one factor while fixing the other two parameters. For example, to evaluate the effect of the number of users on the running time of the algorithm, we fix the number of items and preferences in the training dataset, while varying the number of users by randomly selecting from 20% to 100% of all available users. Then, we compute the average running time for each recommendation. Similar steps have been followed for evaluating the effect of the number of pairwise preferences and the number of items on the running time. The results are presented in Fig.9.

As shown in Figure. 9, the computational complexity is almost constant while varying $M$ and $S$, but, there is a quadratic rise of the running time when the number of items increases.

Note that GRank's computational complexity is quadratic to the number of items since it considers all the possible pairwise comparisons. However, the real number of pairwise comparisons is much less than $N^2$ in many practical applications, because items usually form clusters and the comparisons often happen within those clusters. This means that in practice, we may be able to prune TPG by omitting those comparisons that have not been done by any user.

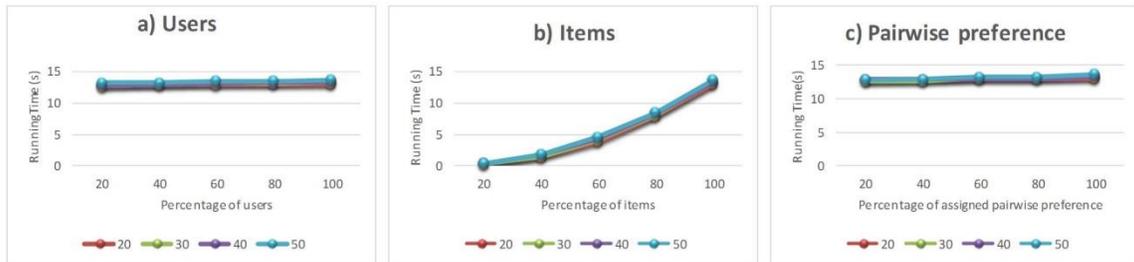

**Fig. 9.** Scalability analysis of GRank in terms of the number of users, items, and pairwise preferences assigned by the users

## 5. Discussion

GRank framework was introduced to resolve the sparsity problem of NCR techniques for similarity calculation. In the following, we briefly discuss how GRank is accomplished to resolve the issues.

- Using TPG, GRank implicitly aggregates different kinds of users' similarities: One type of user similarities is calculated based on their common comparisons. This type of similarity is reflected by the paths following $< U - P - U >$ that connect two users through a pairwise comparison's node. Additionally, two users are assumed to be similar if they both have preferred a particular item A in some comparisons, even if the items over which it has been preferred are different for those two users. The same holds for situations in which two users prefer different items over some particular item A. These types of similarity can be discovered through tracking the paths in form of $< U - P - I_d - P - U >$ and $< U - P - I_u - P - U >$, respectively. Combination and replication of these meta-paths define many different relations among entities and GRank captures and aggregates them in its rank calculation process.

- Following the meta-paths $< P - U - P >$ in TPG, GRank implicitly finds correlated preferences that are the preferences of similar users. Correlate preference are highly connected through paths following $< P - U - P >$. consequently, unknown preferences are iteratively estimated by propagating the known preferences of the target user to those unknown preferences that are correlated to them. This information can be used to calculate users' similarity even in case of no common pairwise comparisons. We refer to an example to clarify the concept: As illustrated in Fig.3, $u_1, u_2$, and $u_5$, each one, has one link to the preference node *B<A, B<D*, and *C<D*. Therefore, the similarity between each pair of them will be zero according to the Kendall correlation. However, TPG reflects that $\{A > B\}$ is highly correlated to the preference node $\{C < D\}$ while it has no relations with B<D. Therefore, $u_1$ is more similar to $u_5$ as a consequence of following the paths passing from $\{A > B\}$ and $\{C > D\}$

- Taking advantages of TPG and PageRank algorithm, GRank directly estimates the users' ranking over unseen items. Personalized PageRank computation in TPG enables GRank to aggregate the ranking information obtained from different forms of meta-paths $< u, v_1, \ldots, v_m, i_d >$ and $< u, v_1, \ldots, v_m, i_u >$ for fast and accurate prediction of users' ranking. Note that this approach differs from the typical 3-step framework (calculation of similarity, generation of preference matrix, and, inference of total ranking) used in all neighbor-based approaches.

## 6. Conclusion

In this paper, we studied how a graph-based framework can be designed and exploited to address the shortcomings of current neighbor-based collaborative ranking algorithms. For this purpose, we suggested that modeling the preference data as a new tri-partite graph structure and then exploring it can help us to capture the different kinds of relations existing in a ranking preference dataset (e.g. users' similarities, items' similarities, etc.). We also proposed a random-walk approach to make recommendation based on the proposed structure. Experimental results showed significant improvement of the suggested framework, GRank over other state-of-the-art graph-based and neighbor-based collaborative ranking methods. It seems that the graph based approach of GRank can be beneficial both in sparse and dense data sets. In dense data sets, it can form the neighborhoods more precisely, by exploring different paths that exist among entities. In sparse data sets, that users rarely have common pairwise comparisons and direct neighborhoods are usually very small, it can still traverse the edges to find farther neighbors and use their information as well for recommendation. The proposed graph structure has been mainly used here for finding closeness between users and items, but it can also be used for other purposes like finding clusters of similar users and similar items, and also discovering correlated preferences which are some of essential concepts in the field of recommendation systems.